\documentclass{JINST}
\newcommand{\arXiv}[1]{\href{http://www.arxiv.org/abs/arXiv:#1}{\tt arXiv/#1}}

\title{HEROICA: an Underground Facility for the Fast Screening of Germanium
Detectors}

\author{E.~Andreotti$^a$,
A.~Garfagnini$^b$$^c$\thanks{Corresponding author.},
W.~Maneschg$^d$,
N.~Barros$^e$, G.~Benato$^f$, R.~Brugnera$^b$$^c$, F.~Costa$^c$,
R.~Falkenstein$^g$, K.~K.~Guthikonda$^f$,
A.~Hegai$^g$, S.~Hemmer$^b$$^c$, M.~Hult$^a$,
K.~J\"{a}nner$^d$, T.~Kihm$^d$, B.~Lehnert$^e$,
H.~Liao$^h$, A.~Lubashevskiy$^d$,
G.~Lutter$^a$, G.~Marissens$^a$, L.~Modenese$^c$,
L.~Pandola$^i$,
M.~Reissfelder$^d$, C.~Sada$^b$$^c$,
M.~Salathe$^d$, C.~Schmitt$^g$,
O.~Schulz$^h$, B.~Schwingenheuer$^d$,
M.~Turcato$^c$, C.~Ur$^c$, K.~von Sturm$^g$,
V.~Wagner$^d$, J.~Westermann$^d$\\
\llap{$^a$}EC-JRC Institute for Reference Materials and Measurements,
Geel, Belgium\\
\llap{$^b$}Dipartimento di Fisica e Astronomia dell{'}Universit{\`a} di Padova,
Padova, Italy\\
\llap{$^c$}INFN  Padova, Padova, Italy\\
\llap{$^d$}Max Planck Institut f{\"u}r Kernphysik, Heidelberg, Germany\\
\llap{$^e$}Institut f{\"u}r Kern- und Teilchenphysik, Technische
Universit{\"a}t Dresden, Dresden, Germany\\
\llap{$^f$}Physik Institut der Universit{\"a}t Z{\"u}rich, Z{\"u}rich,
Switzerland\\
\llap{$^g$}Physikalisches Institut, Eberhard Karls Universit{\"a}t
T{\"u}bingen, T\"{u}bingen, Germany\\
\llap{$^h$}Max-Planck-Institut f{\"ur} Physik, M{\"u}nchen, Germany\\
\llap{$^i$}INFN Laboratori Nazionali del Gran Sasso LNGS, Assergi, Italy}

\abstract{An infrastructure to characterize germanium detectors
has been designed and constructed at the HADES
Underground Research Laboratory, located in Mol (Belgium).
Thanks to the 223~m overburden of clay and sand,
the muon flux is lowered by four orders of magnitude.
This natural shield minimizes the
exposure of radio-pure germanium material to cosmic radiation
resulting in a significant suppression of cosmogenic activation in the
germanium detectors.
The project has been strongly motivated by a special production of germanium
detectors for the GERDA experiment. GERDA, currently collecting data at the
Laboratori Nazionali del Gran Sasso of INFN,
is searching for the neutrinoless
double beta decay of $^{76}$Ge. In the near future, GERDA will increase its
mass and sensitivity by adding new Broad Energy Germanium (BEGe) detectors.
The production of the BEGe detectors is done at Canberra in
Olen (Belgium), located about 30~km from the underground test site.
Therefore, HADES is used both for storage of the crystals over night, during
diode production, and for the characterization measurements.
A full quality control chain has been setup and tested on the first
seven prototype detectors delivered by the manufacturer
at the beginning of 2012.
The screening capabilities demonstrate
that the installed setup fulfills a fast and complete set of measurements
on the diodes and it can be seen as a general test facility
for the fast screening of high purity germanium detectors.
The results are of major importance for a future massive production
and characterization chain of germanium diodes foreseen for a possible
next generation 1-tonne double beta decay experiment with $^{76}$Ge.}
\keywords{Cryogenic Detectors; Gamma Detectors; HPGe Detectors}

\begin{document}

\section{Introduction}
\label{sec:intro}
The GERDA~\cite{bib:GERDA_proposal}-\cite{bib:GERDA_nim}
(GERmanium Detector Array) experiment, in operation
at Laboratori Nazionali del Gran Sasso (LNGS) of INFN
has been designed to search
for the neutrinoless double beta decay ($0\nu\beta\beta$)
of $^{76}$Ge using
germanium detectors. The diodes are made of isotopically modified material,
enriched to about 88\% in $^{76}$Ge, and are operated without
encapsulation in a Liquid Argon (LAr) cryogenic bath.
The experiment is currently running its Phase I
with coaxial germanium detectors 
which were used by the former HdM~\cite{bib:HdM}
and IGEX~\cite{bib:IGEX} double beta decay experiments.
Further information on the construction and commissioning of
GERDA, its physics goals and the first results
can be found elsewhere~\cite{bib:GERDA_nim,bib:GERDA_2nubb}.

While GERDA Phase I is running, the collaboration is preparing a new
phase to double the active detector mass and, with the help of new
LAr veto techniques, to reduce further the overall background aiming at
a higher signal sensitivity.
For the production of new detectors, 35~kg of germanium material
equally enriched in $^{76}$Ge, are available for the GERDA experiment.
Broad Energy Germanium detectors (BEGe)~\cite{bib:BEGe} have been largely
studied for their excellent energy resolution and their enhanced pulse shape
properties which allow to reduce further the background discriminating
multiple from single site events~\cite{bib:bege_pulse_shape}.
Both GERDA and MAJORANA~\cite{bib:MAJORANA}
collaborations have selected similar
detector technologies for their physics goals.
Concerning GERDA, the crystal growth has been performed at Canberra
Oak Ridge (USA)~\cite{bib:canberra}, while the diode production was performed
at Canberra, Olen (Belgium)~\cite{bib:canberra}.

In recent years, a pilot run with isotopically modified germanium
detectors (i.e. depleted in the $^{76}$Ge content) has been performed by
the GERDA collaboration. The procedure allowed to test the production of
commercially available germanium detectors using custom material
and to define the quality acceptance protocol for the
experiment~\cite{bib:GERDA_deplBEGe}.

Special efforts in terms of logistics were taken during the different
production phases in order to minimize the exposure of the enriched germanium
material to cosmic radiation. Cos\-mo\-ge\-ni\-cally produced
radio-isotopes include $^{68}$Ge and $^{60}$Co
that become a serious hazard by mimicking neutrinoless double
beta decay of $^{76}$Ge in germanium detectors. 
The low contamination requirement has been
assured by storing the crystals in a container equipped with shielding
layers of steel and water during transport between the USA and Europe.
Moreover, the crystal first, and the diodes afterwards, are always stored in
underground locations close to the production plants.

The search for a clean underground site where to perform the full set of
characterization measurements on the newly produced diodes before shipping
them to their final destination at the LNGS,
has been successful with the
selection of the HADES (High Activity Disposal Experimental Site)
Underground Research Laboratory. The place is as close as
possible to the diode manufacturer ($\sim 30~\mbox{km}$), and besides hosting
the characterization measurements, it is used to store the crystals and the
diodes during production over night and during weekends.

The present paper starts with a description of the unique characteristics
of HADES as a low background site for
detector testing and low activity measurements.
After a short introduction to the diodes acceptance protocol, the
infrastructure (electronics, computing and DAQ resources),
the characterization setups and 
their performances are described.
Furthermore, the overall performances of the acceptance tests
are reviewed and the prospects for the characterization of the full batch of
detectors to be produced for GERDA are given.
%

\section{The HADES Underground Research Laboratory}
\label{sec:hades}
\begin{figure}[t]
\centering
\includegraphics[width=0.95\textwidth]{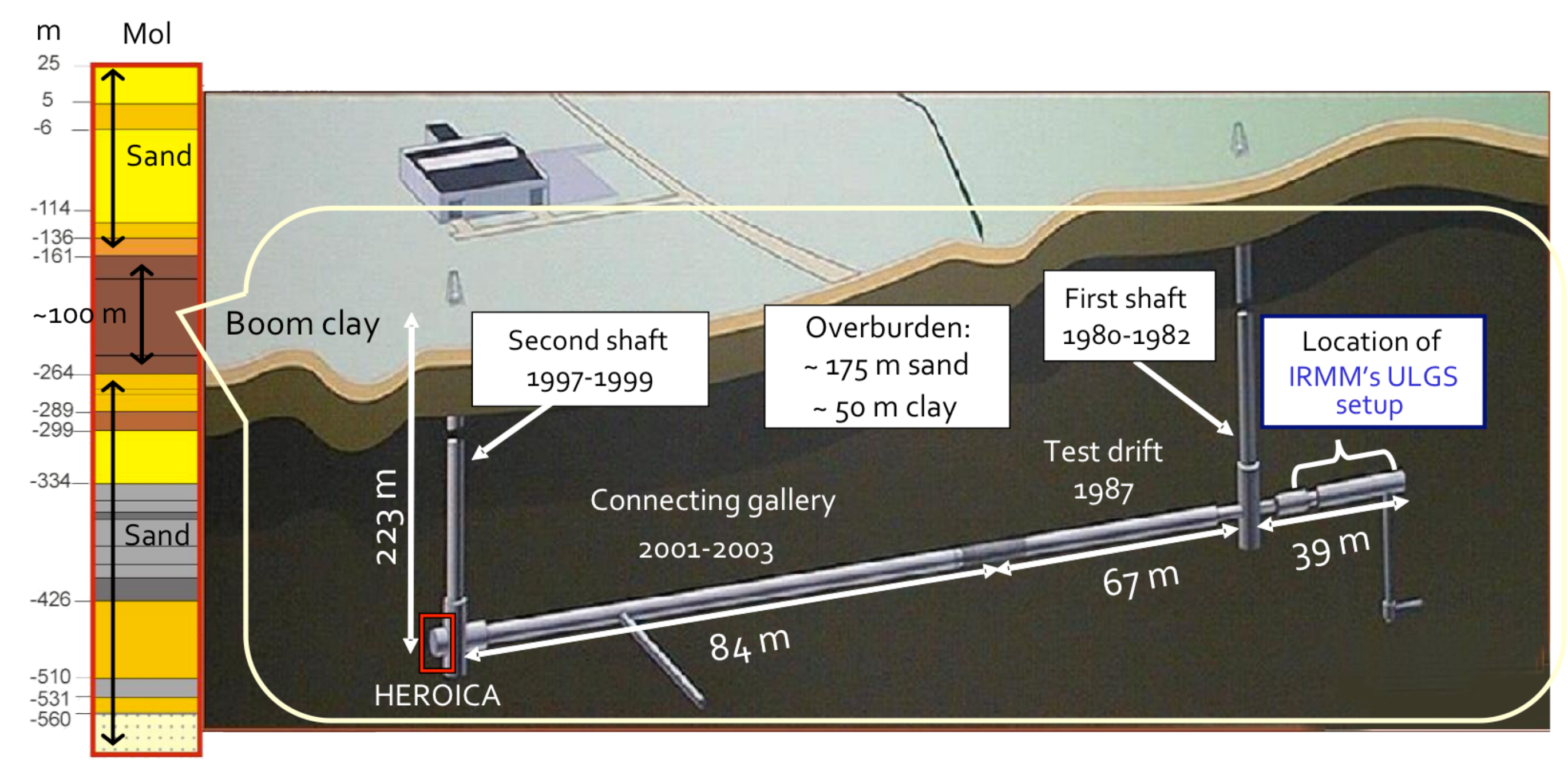}
  \caption{\label{fig:hades:layout}
Layout and construction history of the underground laboratory at
HADES. The HEROICA test facility is located behind the second
shaft (bottom left corner). On the right-end side of the tunnel, the
ULGS facility is hosted. Further discussions and references are given in the
text. From~\cite{bib:hades}.}
\end{figure}

HADES is a semi-deep underground laboratory~\cite{bib:hades}
built in the frame of the Belgian research programme regarding
the geological disposal of radioactive waste on the premises of the
Belgian Nuclear Research Center SCK~$\cdot$~CEN (Studie Centrum voor
Kernenergie $\cdot$ Centre d'Etude de l'Energie Nucl\'eaire) in Mol (Belgium).
Since 1997 HADES is managed by EURIDICE (European Underground Research
Infrastructure for Disposal of nuclear waste In Clay Environment) as
a research facility~\cite{bib:hades_scientific_programme}
and not as a disposal site.
As shown in Figure~\ref{fig:hades:layout}, the underground laboratory
is located at about $223~\mbox{m}$ below ground in the "Boom clay" layer.
The main purpose of the
laboratory is to examine the possibility of constructing a high level
nuclear waste repository at such a depth and conduct various in-situ
experiments with the aim of characterizing geological properties of the clay.
The underground tunnel is reachable via two shafts.
The Institute for Reference Material and
Measurements (IRMM), part of the European Commission's Joint Research Centre
(JRC), operates a dedicated Ultra Low-level Gamma-ray Spectrometry (ULGS)
facility at one end of the HADES tunnel~\cite{bib:andreotti_ulgs}.

An area of about $15~\mbox{m}^2$ at the opposite end of the gallery
has been assigned to the HEROICA
(Hades Experimental Research Of Intrinsic Crystal Appliances) project.
The area has been equipped with dedicated setups, DAQ systems and networking
for data transfer to outside institutes.
In order to minimize microphony on the highly sensitive devices a new
damped pavement (10~mm thick) has been installed.
Moreover, all electronic devices have been grounded in order to 
minimize noise and cross-talk effects.
%

The sand and clay overburden corresponds to about 500~m water equivalent
and cuts the muon flux and secondary hadronic showers down to
 $10^{-1}\mbox{m}^{-2}\mbox{s}^{-1}$ which is
a reduction of about four orders of magnitudes with respect to the ground
level~\cite{bib:andreotti_ulgs}.
The radon concentration in the HADES tunnel is low; the average value collected
over a fortnight test measurement using a Radim 3A~\cite{bib:radim3a}
detector is $21~\mbox{Bq}/\mbox{m}^3$, with oscillations between
$6$ and $43~\mbox{Bq}/\mbox{m}^3$.
Such a low radon level is due to low uranium concentration
in the surrounding material and to a constant ventilation of the
underground areas with an air flux of about $10000~\mbox{m}^{3}/\mbox{h}$.
Even though the HEROICA detector shielding does not include an active
anti coincidence muon veto and nitrogen flushing in order to further
minimize the impact of radon (and its daughters), the obtained background
is highly satisfactory in terms of the HEROICA acceptance tests
and for storage of the enriched germanium material.
Indeed, the HEROICA setups are designed for the characterization of germanium
detectors using strong calibration sources.
On the contrary, low-level activity IRMM
detectors at ULGS site are designed to measure low-activities
in material samples and thus require the best achievable background
suppression applying different shielding techniques.


\section{The Germanium Detector Specifications and Measurement Protocol}
\label{sec:measproc}
The acceptance tests aim to verify the specifications
given by the manufacturer, to extract important detector parameters
(such as the active mass and dead layer thickness) and to
determine the optimal operational conditions.
Since the detectors will be operated "naked" in an underground experiment,
this is the only chance to fully characterize the detectors before
dismounting them from their vacuum cryostat and deploy them
in the GERDA LAr cryostat (as described in~\cite{bib:GERDA_nim})

As mentioned above, a dedicated run with isotopically modified germanium
detectors, depleted in the $^{76}$Ge content, has been very important
to optimize the acceptance test protocol~\cite{bib:GERDA_deplBEGe}.

The operational parameters to be determined are:
\begin{itemize}
\item depletion voltage;
\item energy resolution;
\item leakage current.
\end{itemize}

The test is performed to study the charge collection features of each diode and
determine
\begin{itemize}
\item the detector active volume and mass;
\item the dead layer thickness and uniformity over the surface.
\end{itemize}
Finally, the pulse shape characteristics of the diodes have to be studied and
their efficiency for identification of single-site events ($0\nu\beta\beta$
candidates) and rejection of multiple-site events (background) has to be
determined:
the study allows to refine background reduction techniques for
double beta decay searches of $^{76}$Ge~\cite{bib:GERDA_nim},
\cite{bib:HEROICA_7prototypes}.

According to the contracts and agreements between the GERDA collaboration
and Canberra Industries, the new detectors are
delivered to HADES mounted in a vacuum cryostat and
assembled in the Canberra Dip-Stick vertical dewar~\cite{bib:pst:7500_SL}.
The dewar is positioned on  a moving cart which allows to move the detectors
within the test area and to position them on the measurement stands.

The BEGe specifications are the following:
\begin{itemize}
\item crystal diameter $75 \pm 5~\mbox{mm}$;
\item crystal length $30 ^{+10}_{-5}~\mbox{mm}$,
      potentially up to $50~\mbox{mm}$;
\item detector mass $>600~\mbox{g}$;
\item energy resolution (FWHM) better than 2.3~keV at
      the 1332~keV $^{60}$Co peak;
\item leakage current below $50~\mbox{pA}$;
\item full depletion voltage below 4~kV.
\end{itemize}

%
%
%
%
%

\section{The HEROICA Project}
\label{sec:heroica}

Two different mechanical setups have been designed for the tests of the diodes:
\begin{itemize}
\item a measurement table which allows: 1) to shield the diode
      from the environmental background and protect the operators from
      the radiation created by the simultaneous use of several calibration
      sources in the same area; 2) to place the calibration sources
      (with or without collimators) in a few fixed positions
      around the detector;
\item a setup with a movable arm, motor controlled, which allows to perform
      a top and lateral detector surface scans with a collimated source.
\end{itemize}
In the following sections both measurement setups will be described in detail.

\begin{figure}
 \begin{minipage}[b]{.45\linewidth}
  \includegraphics[scale=0.085]{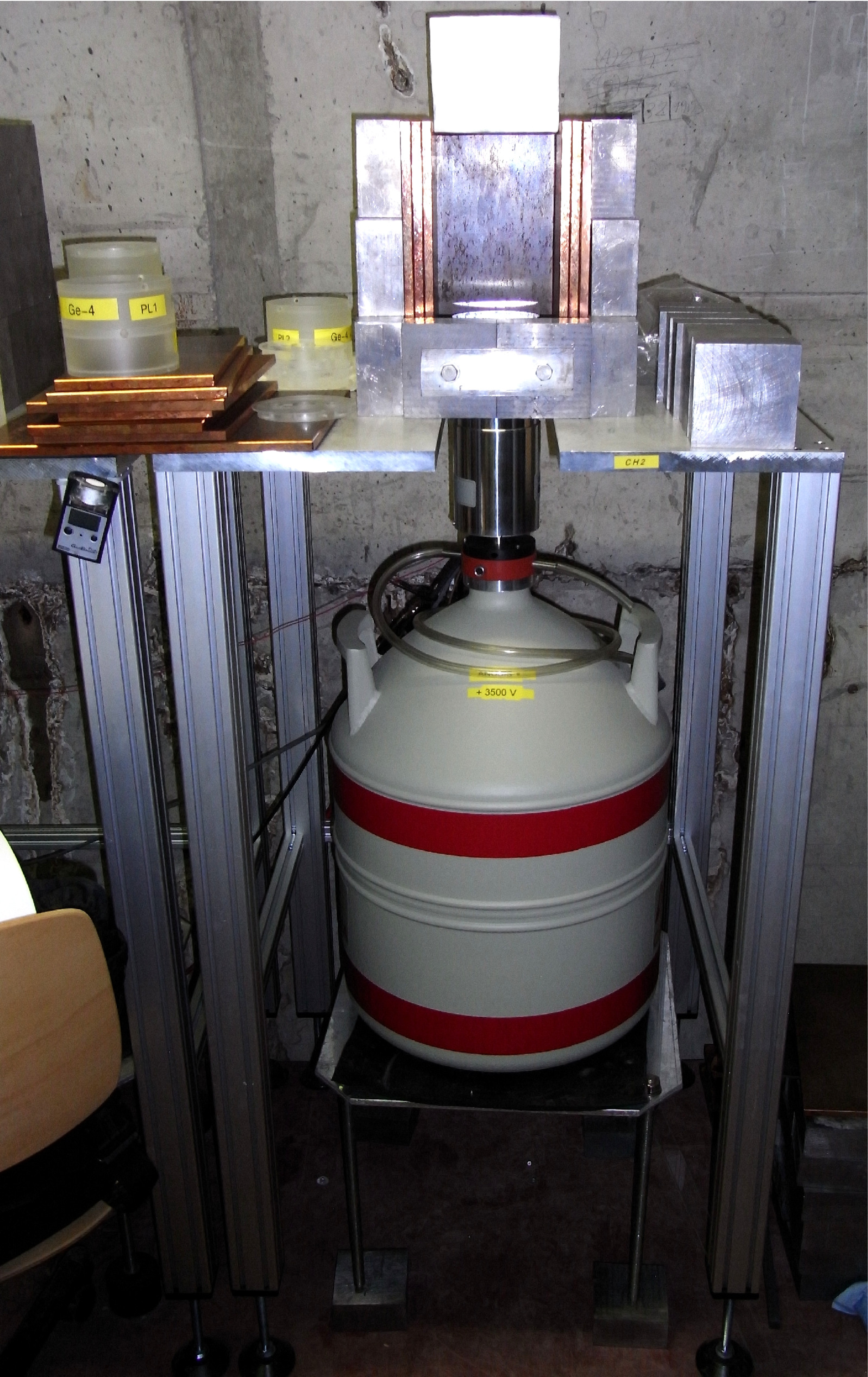}
 \end{minipage}
 \hspace{.05\linewidth}
 \begin{minipage}[b]{.45\linewidth}
        \includegraphics[scale=0.13]{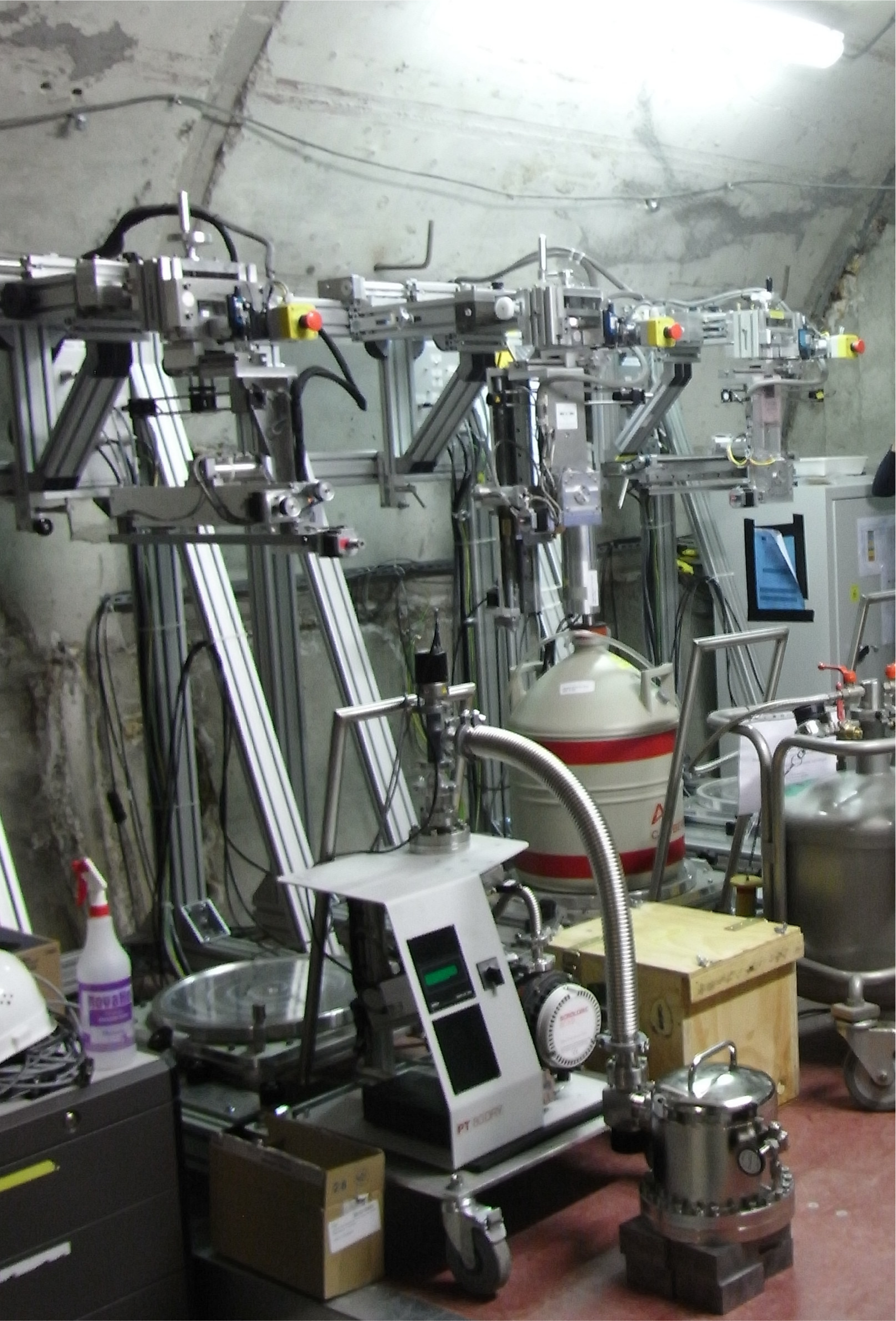}
 \end{minipage}
\caption{\label{fig:setups} Left: One of two setups designed for static
source measurements.
Right: Three automated scanning tables for dynamic source measurements.} 
\end{figure}

\subsection{The Fixed Calibration Measurement Setup}
\label{sec:heroica:hd_table}

The characterization of several standard parameters describing
the overall performance of germanium detectors such as depletion voltage,
energy resolution, average dead-layer thickness and pulse shape performance
can be done with sources positioned at a fixed distance from the detector.
For this purpose, the HEROICA infrastructure includes two setups that consist
of a table with appropriate shielding and source holders.
The tables have adjustable height and a bay which allows to park a movable
cart carrying a cryogenic dewar and the
dip-stick cryostat at the center of the setup
(see Figure~\ref{fig:setups}, left).

The passive shield is built from $10 \times 10 \times 5~\mbox{cm}^3$
lead bricks, so that the shield thickness is 5 cm. Three additional layers
of copper plates (each 1~cm thick) are inserted in the inner part of the
shielding castle.
The shielding is completed with a 1 cm copper plate on the top.
A cross section of the shielding castle can be seen in
the left picture of Figure~\ref{fig:setups}.

Thanks to the shielding,
the exposure of shifters to radiation by a large set of sources
used during the long-standing screening activities is significantly reduced.
The height of the shielding allows to position sources at distances
up to about 20~cm above the detector.
As discussed previously, further background suppression by an
active muon veto or
a radon tight, nitrogen flushed detector chamber is not needed.

\subsection{The Automated Scanning System}
\label{sec:heroica:pst}
An automated scanning table has been designed and built to move a collimated
source around the detector cryostat.
The setup allows to irradiate the detector top and lateral surfaces.
All measurements are performed with a 5~MBq $^{241}$Am source
(see Table~\ref{tab:radsources} for a list of the calibration sources)
placed inside a shielding box ($30 \times 30 \times 30~\mbox{mm}^3$), made
of copper with a 1~mm collimator hole.
The purpose of the measurements is the study of the charge collection
efficiency of the diode by looking at the detector response to the source
placed on different positions on the top and lateral surfaces.
Thanks to the 59.5~keV $^{241}$Am photons, which are easily absorbed
by the collimator walls, a pencil-like beam is shot on the cryostat
and it allows to estimate the diode dimensions.

The collimated source is mounted on an arm which is parallel to the
diode top surface. The source can be moved along the arm (i.e. along the
top surface of the diode) with a positioning resolution and reproducibility
better than 1~mm.
Moreover, the arm can be rotated around the crystal cylindrical axis between
0 and 359 degrees in steps of 1 degree.
\begin{figure}
\centering
\includegraphics[width=0.95\textwidth]{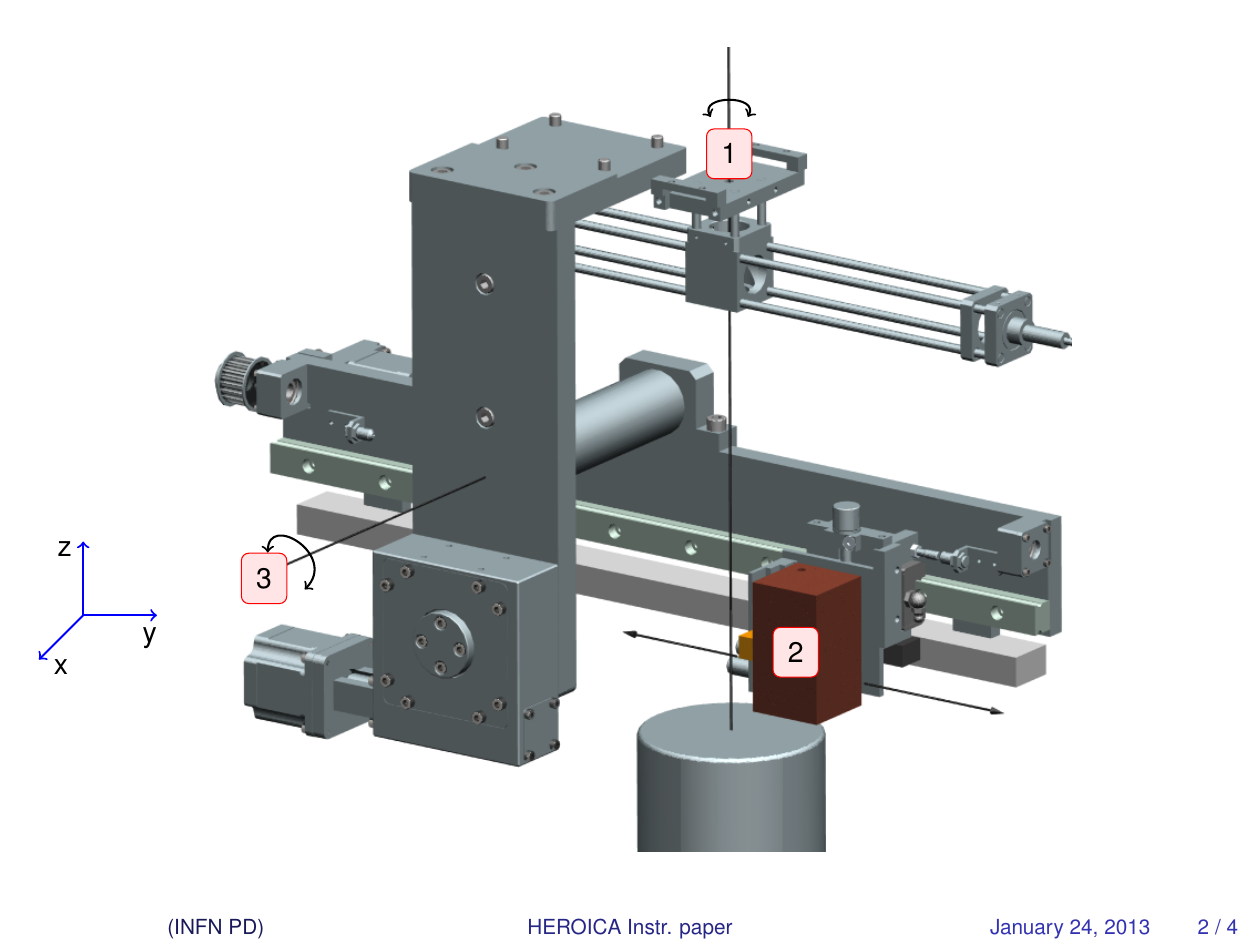}
%
%
%
%
%
%
%
%
%
%
%
  \caption{\label{fig:pst:basic_movements}
Automated Scanning Table: detailed drawing of the possible movements:
1) rotation of the full system around z-axis (the diode cylindrical axis);
2) source holder movement;
3) rotation of the main arm, holding the source.
Further descriptions are given in the text.}
\end{figure}

Figure~\ref{fig:pst:basic_movements} shows a detail of the setup with
special emphasis on the main movements. Label 1 shows the possible rotation of
the full system around the detector z-axis. The source, mounted inside a
copper collimator (label 2) can be moved along the arm holder. By acting
on movements, labelled (1) and (2) on Figure~\ref{fig:pst:basic_movements},
it is possible to reach any point on the x-y plane.
Finally, movement (3) allows to rotate the source holding arm around the x-axis,
thus bringing the arm along the z-axis.
In this new position, by acting on movements
1 and 2 it is possible to move the source around the detector lateral
surface.

%
%
\begin{figure}
\centering
\includegraphics[width=0.95\textwidth]{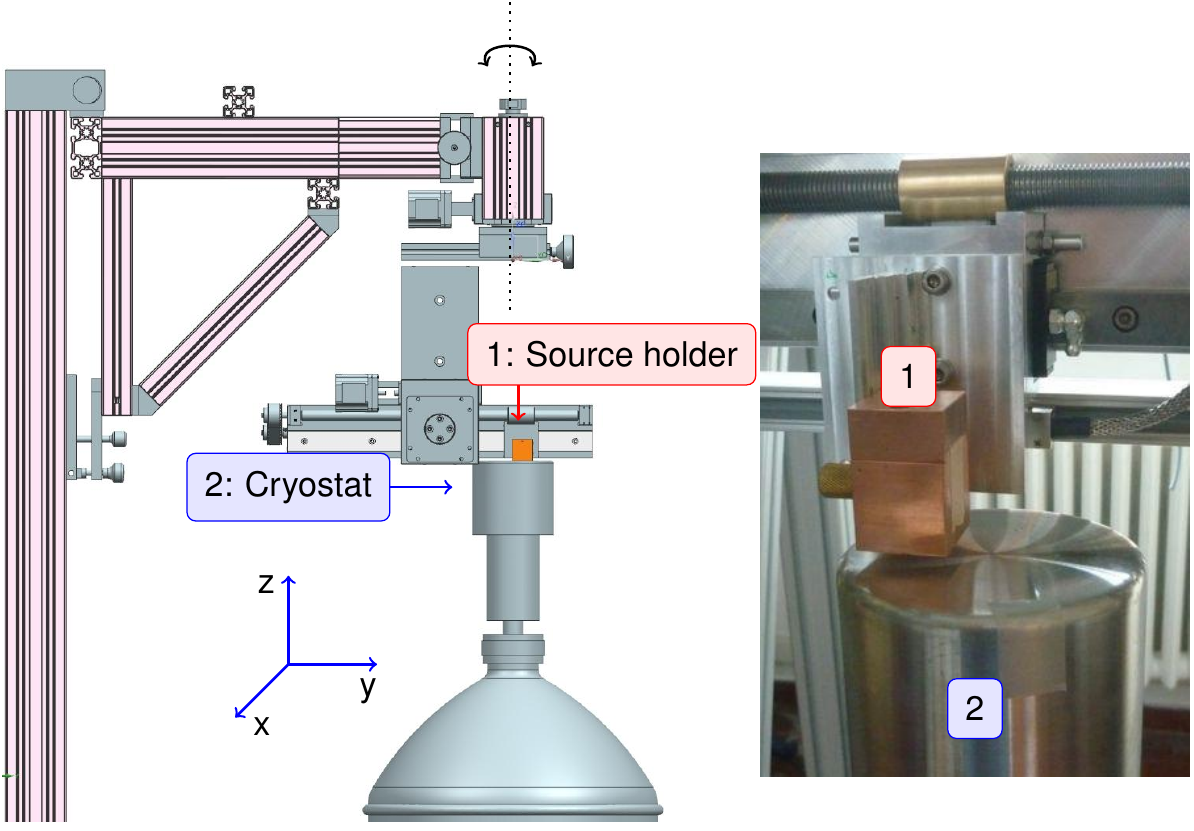}
  \caption{\label{fig:pst:setup_draw_top}
Automated Scanning Table: detector top surface scan.
Left: drawing of the mechanics and source collimator during a top scan.
Right: picture of the copper collimator holding the source during a top scan.
The source holder (label 1) and the cryostat (label 2)
are explicitly indicated in both figures.}
\end{figure}
\begin{figure}
\centering
\includegraphics[width=0.95\textwidth]{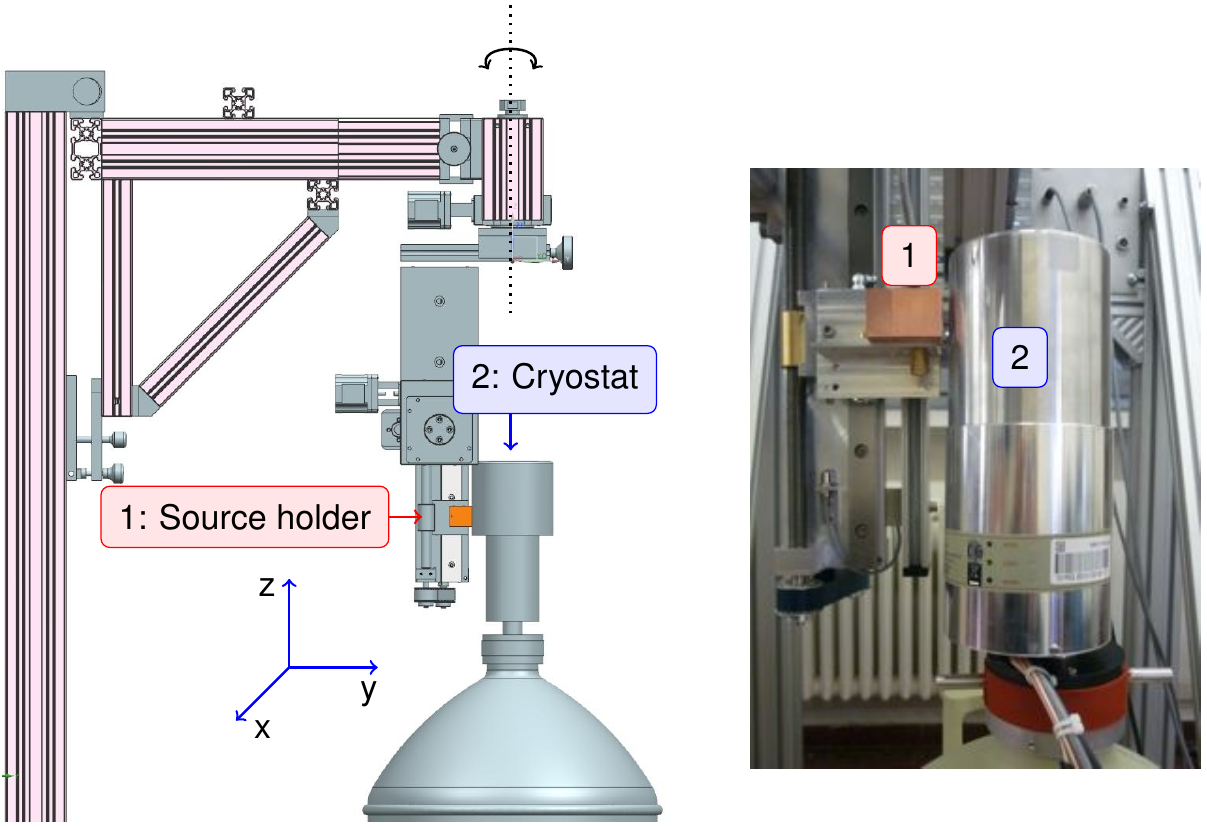}
  \caption{\label{fig:pst:setup_draw_lat}
Automated Scanning Table: detector lateral surface scan.
Left: drawing of the mechanics and source collimator during a lateral scan.
Right: picture of the copper collimator holding the source during a lateral
scan.
The source holder, Cu absorber, (label 1) and the cryostat with the diode
(label 2) are explicitly indicated in both figures.}
\end{figure}

Figures~\ref{fig:pst:setup_draw_top} and \ref{fig:pst:setup_draw_lat}
show the two main apparatus settings while scanning
top and lateral surfaces, respectively.
In Figure~\ref{fig:pst:setup_draw_top}, the arm is placed horizontally
in the x-y plane, parallel to the top detector surface, and the source holder is
positioned on top of the cryostat.
While keeping the arm fixed, it is possible to move the source along it,
thus performing a scan along a top surface diameter.
Since the arm can be rotated around the z-axis (which coincides with the
diode axis), it is possible to span various diameters.
Moreover, by keeping the source still with respect to the arm, while rotating
it around the z-axis, circumferences are spanned on the top surface.
The right part of Figure~\ref{fig:pst:setup_draw_top} shows in detail
the top surface cryostat with the source holder on top of it while
a measurement is running.

In Figure~\ref{fig:pst:setup_draw_lat}, the source holding arm has been
rotated by 90~degree around the x-axis (movement 3 in
Figure~\ref{fig:pst:basic_movements}). The source can be moved along the
z-axis, while the full system can be rotated in the x-y plane,
around the z-axis.

\subsection{Radioactive Sources and Source Holders}
\label{sec:rsources}
A large set of radioactive sources has been provided according to the
requirements of the different measurements included in the test protocol.
This large number is due to the need to perform a fast screening on a large
number of detectors (up to 5 detectors are measured in parallel with a global
rate of two diodes fully screened per week).
The choice of the source isotope and activity for each measurement is
driven by the detector's properties to be investigated and the
acceptable statistical uncertainty of the measurements; the latter is a
compromise between DAQ trigger and filter limitations, required measurement
time and number of collected events.

Table~\ref{tab:radsources} shows a list of commonly used sources for the
standard characterization measurements.

\begin{table}[htp]
\centering
\begin{tabular}{|c|c|c|c|} \hline
amount   & nuclide  & activities  &  measurements \\
         &          & [kBq]       &  \\ \hline
 5       & $^{60}$Co & 3-15       &   Resolution, High Voltage Scan,
                                     Active Volume \\
 3       & $^{228}$Th & 10        & Pulse Shape Analysis \\
 5       & $^{241}$Am & 100-500   & Dead Layer, Pulse Shape Analysis \\
 5       & $^{133}$Ba & 3-45      & Dead Layer \\
 3       & $^{241}$Am & 5000      & Charge Collection \\
 3       & $^{137}$Cs & 5-20      & Energy Calibration  \\
 1       & $^{152}$Eu & 80        & Energy Calibration  \\ \hline
\end{tabular}
\caption{List of the most commonly used radioactive sources in
the HEROICA experimental area for BEGe characterizations.
In the list are included the three high
activity $^{241}$Am sources, about 5~MBq, used in the automated scanning
tables.}
\label{tab:radsources}
\end{table}

For the automated scanning measurements, a high count rate is required
so that even with a collimated source the total net count in the
59.5~keV peak of $^{241}$Am is of the order of 10$^{3}$ when measuring for
less then 5 minutes in each point.
Therefore, an activity of 5 MBq was chosen.
For the high voltage scan measurements, a $^{60}$Co source of
$\sim 10~\mbox{kBq}$ was chosen, which allows to obtain
$\sim 10^{3}$ net counts in the two main peaks at 1.1~MeV and 1.3~MeV
within a 10 minutes
measurement (at each high voltage value).
In this way, a complete high voltage measurement lasts less then one day.
For the active volume determination, a $^{60}$Co source is used, while
for the dead layer determination, a $^{241}$Am and a $^{133}$Ba source are used.
For these measurements, a net count of more than $10^{4}~\mbox{events}$
in the peaks of interest are required within 1-2 hours.
A $^{228}$Th source of $\sim 10~\mbox{kBq}$
is used for the pulse shape analysis,
which allows to obtain a statistics of about $10^{4}$ events in the
double escape peak (at 1.59~MeV)
of the $^{208}$Tl 2.6 MeV line with a measurement time of about 8 hours.

For all static measurements, dedicated plexiglass source holders have been
procured. They consist of a base, fitting on the cryostat cap so that it is
properly centered, and on which it is possible to insert one or a series of
different plexiglass pieces with fixed height. In this way, each
measurement can be performed with a fixed and reproducible
source-to-detector distance.
%

\subsection{Front-End Electronics and Data Acquisition}
\label{sec:heroica:electronics}
The detectors under test are connected to three two channel ISEG~\cite{bib:iseg}
HV power supplies (mod. 246L), operated and controlled
via CANbus~\cite{bib:canbus} interface.
The front-end read out is performed with Canberra 2002 CSL
charge sensitive preamplifier with RC-feedback
preamplifiers (decay time $\tau = RC \sim 47\mu\mbox{s}$) including cold FETs.
A pulser signal can be fed into a test-input line of the preamplifiers for
checking the signal stability over time.
Another test-voltage point connected to the preamplifier allows to monitor
continuously the leakage current which is subsequently registered by voltage
loggers EL-USB-3 from LASCAR \cite{bib:loggers}.
The output signals can be read by two DAQ systems:
Multi Channel Analyzer modules by ORTEC and Canberra,
and Struck Flash ADCs.

%
Two different MCA modules are used for the characterization
measurements:
\begin{itemize}
\item the Canberra~\cite{bib:canberra:mca}
Multiport II NIM module for the fixed calibration
measurement setup (up to six independent inputs are available);
\item the ORTEC~\cite{bib:ortec:mca}
926 (single buffer) and 927 (dual buffer) for the
measurements done with the automated scanning tables.
\end{itemize}

%
The full-registration of the waveforms of single events is performed with
two Struck SIS3301 VME FADCs and  2~GHz VME CPUs that are mounted into
a Wiener VME crate.
Each FADC module accommodates up to eight input channels allowing
a sampling-rate of 100~MHz.
A pulse is saved with 14-bit resolution, such that it contains up to
128~k samples with a maximum trace-length of 1.28~ms.
In order to make full use of the dynamic range, a custom-made amplifier
has been installed.
Besides the ability of sampling, the Struck FADC allows to apply fast
integration and differentiation filters to shape signals and produce
energy spectra. Moreover, the FADC works in a dual buffer mode, i.e.
the non-active buffer can be read out during data-collection.
However, the Struck FADC does not include the synchronization of modules
operating in parallel, a usable clock for absolute time estimation of stored
traces, and a reliable implementation of a counter used for a correct live time
calculation.
These tasks are performed by a Field Programmable Logic Array on
a custom made VME module~\cite{bib:GERDA_nim}.

\subsection{Electronic Noise Optimization}
\label{sec:noise-red}
In order to have an optimal energy resolution and
amplitude versus energy (A/E)
distribution\footnote{The amplitude versus energy (A/E) distribution
is used for pulse shape discrimination techniques. $A$ is the maximum height
of the differentiated input signal, while $E$ is the corresponding
reconstructed energy.}
for the BEGe detectors under test, it is essential to
operate in a noise-free environment.
As mentioned previously, a first step in this direction was the
installation of a custom-made 10\,mm thick damped pavement to reduce
microphonic noise originating from the elevator close to the HEROICA site
and from the stainless steel floor and its
support structure inside the HADES tunnel.
In addition, after the installation of the static source tables and
of the first automated scanning table, dedicated noise studies were performed.
In the first version of the setup, a periodically modulated signal
originating from the scanning table, was detected.
%
During the installation of the second and third scanning tables,
a full grounding of all setups and electronics was performed,
which significally improved the overall noise.
The results are discussed elsewhere~\cite{bib:HEROICA_7prototypes}.


\subsection{Networking and Computing Infrastructures}
\label{sec:heroica:computing}
A dedicated network (virtual LAN) has been setup for the computing facilities
of the project. A public server (bastion host type) allows to access the
internal resources from the outside and provides all network services
required by the internal network nodes (DNS, DHCP). The access from/to the
Internet is restricted to specific hosts and services and is handled
through a firewall administered by the SCK$\cdot$CEN, IT group.
A limited number of
hosts, one for each participating institute, has been granted authorization
for ssh~\cite{bib:comp:ssh} and
VNC~\cite{bib:comp:vnc} access to the bastion host.
Due to the access restrictions\footnote{Access to the area is possible
during working days from 7:30 to 16:00 and always supervised by SCK
authorized personnel.}  to the HEROICA experimental area,
remote access is crucial for
controlling the status of the tests and for the real-time check
of the collected data.
All data are automatically copied to computing resources hosted 
at MPIK Heidelberg, during the night. The procedure, besides providing a
backup copy of all the data, allows to perform remote analyses.

Concerning the computer infrastructure:
\begin{itemize}
\item a bastion host 
running LINUX Debian 6.0
which acts as a disk server for all data
(an array of RAID'5 disks for a total of 14 TBytes), has a local
database (PostgreSQL) for monitoring of detectors high voltages, and is used
to control the data acquisition;
\item a virtual machine 
running Windows XP
which is used to control the scanning tables and
to run the Canberra and ORTEC MCA.
\item a VME CPU, running LINUX Ubuntu 8.10, used to control the Struck
FADC modules.
\end{itemize}

\section{Screening Capacity and Performances}
\label{sec:results}
According to the screening protocol, after a new detector is delivered,
the following measurements are performed on a static table:
\begin{itemize}
\item energy resolution; a run with $^{60}$Co at the operational voltage
      suggested by the producer is taken.
      The measurement runs for 10 minutes with MCA and FADC
      histogram mode\footnote{In histogram mode only the reconstructed energy
spectrum is saved on disk, while the single event pulses are discarded.}.
\item $^{60}$Co high voltage scan; several runs at different bias voltages are
      collected and the resolution and peak integral studied as a function
      of the voltages. The single runs last between 2 and 5 minutes and with
      a voltage step of 50~V, the scan will last between 5 and 8 hours.
\item active volume determination;
      a long run (about one hour) is performed with a $^{60}$Co
      source at full depletion voltage.
\item dead layer measurements; two runs are collected using
      uncollimated $^{241}$Am and $^{133}$Ba sources.
      Depending on the source activity,
      the runs last one hour for $^{133}$Ba and at least two hours
      for $^{241}$Am.
\item pulse shape discrimination; a long run with a $^{228}$Th source is
      collected for at least eight hours. The acquisition is performed
      collecting the full pulse waveform for further analysis.
\end{itemize}


In a second phase, the detector is moved to the scanning table and the
diode charge collection is studied using a collimated $^{241}$Am source:
\begin{itemize}
\item a full scan on the top surface along two orthogonal diameters is
      performed. The measurements with MCA data acquisition is taken
      in steps of 1~mm or less and with
      a live time of 2 or 3 minutes. The data are used to estimate the diode
      dimensions
      and verify charge collection uniformity along the surface.
      Moreover the measurements
      are used to align the detector in the analysis reference system.
\item a circular scan on the top surface is taken. The source spans
      two outer circles close to the detector border, two circles in the
      middle of the diode and one point in the centre of the detector.
      The measurements are collected with both MCA and FADC (waveform
      sampling mode) and are used to perform detailed studies
      on the A/E distribution on the detector top surface.
\item one or two linear scanning measurements are performed on the lateral
      surface. The runs are taken with the MCA data acquisition system and
      are collected to verify charge collection uniformity on the lateral
      surface and to check the crystal height.
\item a circular scan on the lateral surface is collected. A ring at a fixed
      height is taken with both MCA and FADC (waveform sampling mode)
      and is used to check the A/E distribution on the lateral surface.
\end{itemize}


%
\begin{figure}
\centering
   \includegraphics[width=0.75\textwidth]{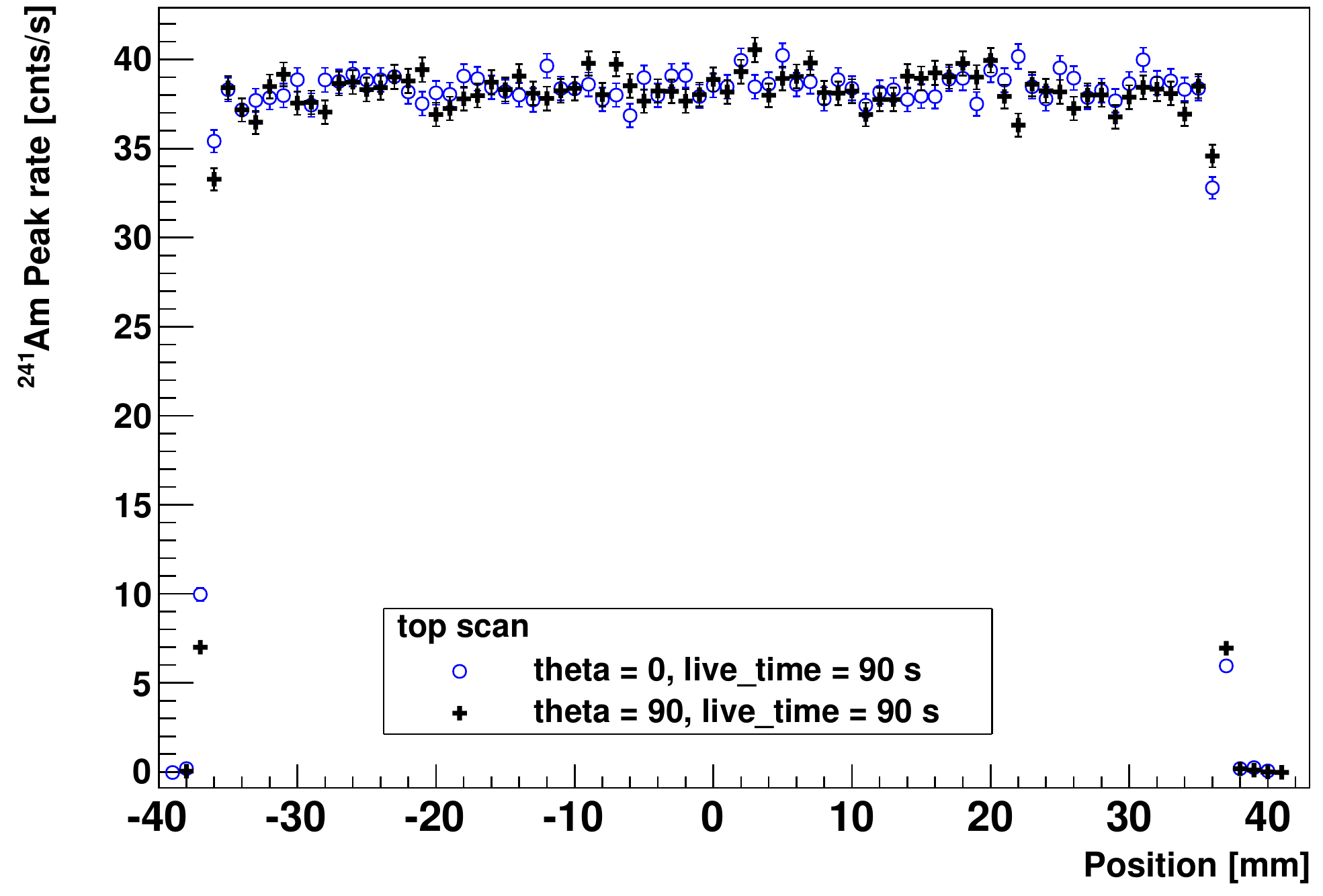}
  \caption{\label{fig:pst:topscan_ge9}
Germanium detector top surface scan. The plot shows the $^{241}$Am
59.5~keV peak count rate as a function of the source position, along
the diode diameter, above the detector surface.
The source is moved in steps of 1~mm between different
measurements. Two different sets of measurements are presented, 0 and
$90^{\circ}$ for the arm position with respect to the detector cylindrical axis.
Further discussion is given in the text.}
\end{figure}

While a complete discussion of the measurements and the results obtained
from the tested diode is the argument of a separate
paper~\cite{bib:HEROICA_7prototypes}, the automated
scanning tables performances are presented in the following.
The setup represent an innovative step compared to the work
presented in~\cite{bib:GERDA_deplBEGe}).

An example of detector response to $^{241}$Am photons during a top
surface scan can be seen in
Figure~\ref{fig:pst:topscan_ge9}: the histogram reports the 59.5~keV peak
count rate as a function of the source position above the
detector surface. The collimator is moved along
a straight line with 1~mm steps above the detector endcap.
Two sets of measurements are presented: for the second set (crosses),
the source arm is rotated by 90~degree in the x-y plane
with respect to the first set (open circles).
The count rate is zero outside the detector active
volume and reaches a constant value of about 40 counts/s while on the diode
surface.
This fast measurement (the exposure is only 90 seconds per point)
allows to determine the active detector diameter.
\begin{figure}
\centering
   \includegraphics[width=0.75\textwidth]{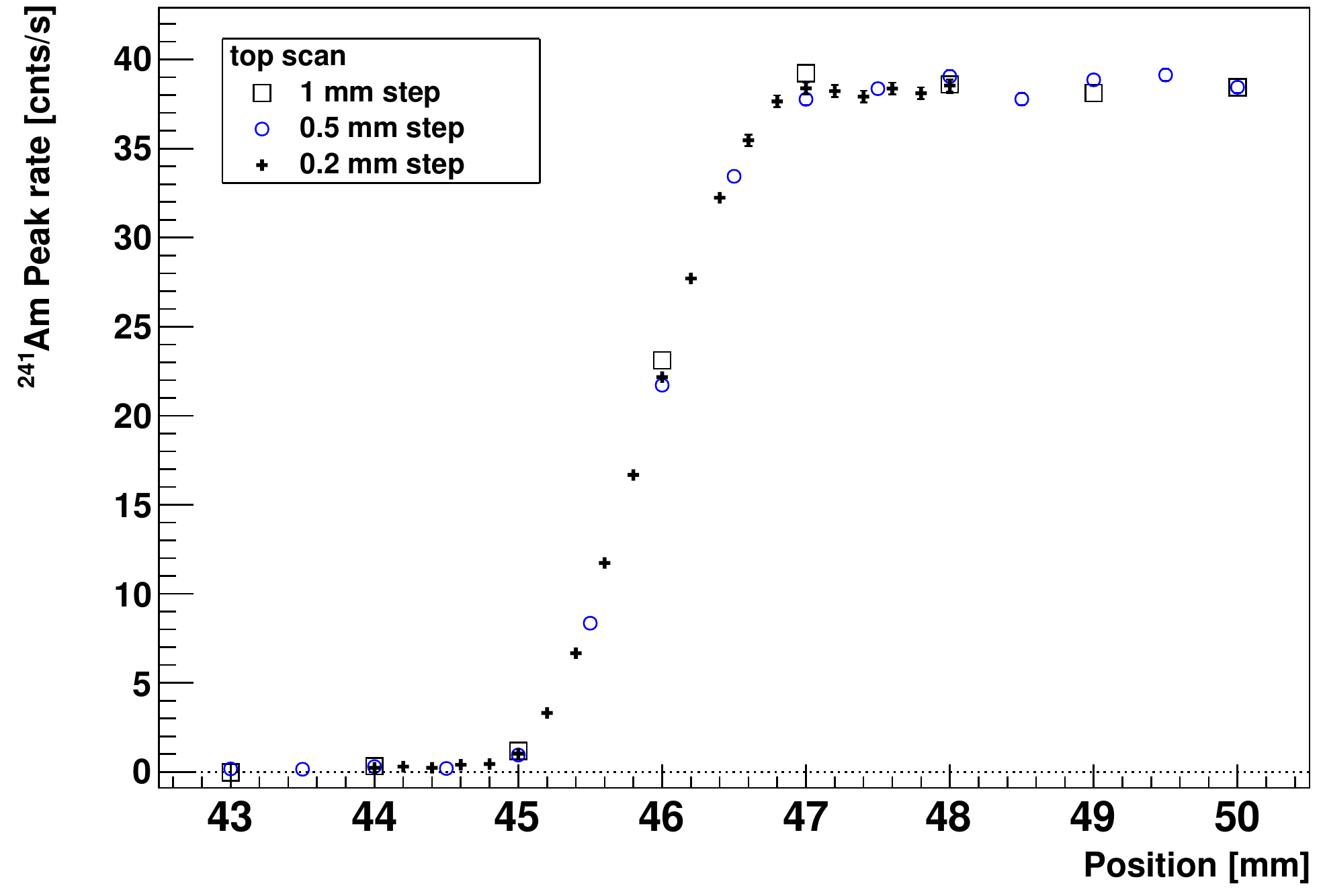}
  \caption{\label{fig:pst:topscan_edge}
Germanium detector top surface edge scan. The same scan is performed three
times with different steps between the points: 1~mm (open squares), 0.5~mm
(open points) and 0.2~mm (crosses). The plot shows a remarkable overlap among
the different points.}
\end{figure}
\begin{figure}
\centering
   \includegraphics[width=0.75\textwidth]{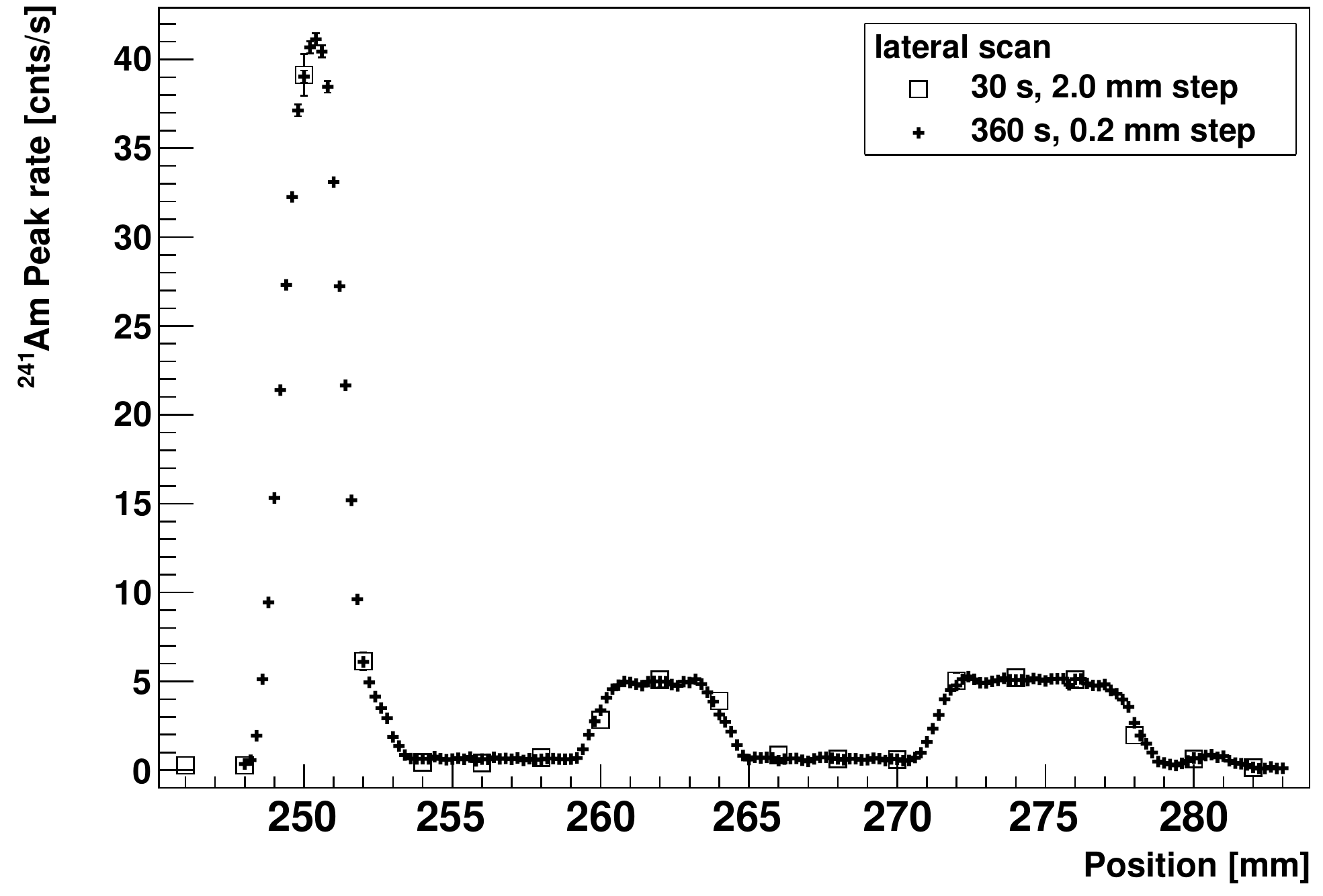}
  \caption{\label{fig:pst:latscan_two}
Germanium detector lateral surface scan. The scan is performed at a fixed
angle with two different step sizes: 2~mm (open squares) and 0.2~mm
(crosses). The points show a very good overlap. The shape of the count
rate is determined by the structure of the holder around the diode; the
suppression in the rates is due to two copper rings in the regions 252-260~mm
and 264-271~mm.}
\end{figure}
A study of the reproducibility of the source positioning on the detector
surface has been performed. Figure~\ref{fig:pst:topscan_edge} shows three
sets of measurements taken on the same detector edge with different
step size among the source positions: 1~mm, 0.5~mm and 0.2~mm.
The integrals of the photon contributing to the $^{241}$Am 59.5~keV peak are in
remarkable agreement between the different sets of measurements.
The source is moved to the same zero position (located outside the detector)
before starting each set of measurements.

The response of the detector to $^{241}$Am photons during a lateral
surface scan can be seen in Figure~\ref{fig:pst:latscan_two}:
the histogram reports the 59.5~keV peak
count rate as a function of the source position along the lateral surface
of the detector. The shape of the count rates is completely determined
by the structure of the detector holder used to support the diode inside
the vacuum cryostat: the strong suppression rate
in the regions 252-260~mm and 264-271~mm is indeed due to two reinforcement
rings present in the holder mechanics. The reduction of measured count rate is
in very good agreement with the holder thickness and geometry provided by the
manufacturer.

\section{Conclusions}
\label{sec:conclusions}
A full infrastructure has been designed, built and commissioned for the
acceptance tests of the new, $^{76}$Ge enriched,
BEGe detectors for the Phase II of the GERDA
experiment. The facility has been run at first for a period of five months,
in early 2012, on the first seven BEGe prototypes delivered by Canberra.
Starting from the
end of August 2012 it is working to test the remaining 23 BEGe detectors with
an average speed of 2 diodes per week.

The infrastructure has demonstrated an high flexibility and has
permitted to collect an enormous amount of information on each tested diode.
These pieces of information will be essential in the future application of the
characterized diodes in the GERDA experiment.

\section*{Acknowledgments}

The authors would like to thank the team of EIG EURIDICE for their
support during the installation phase and during the running of the project.
Many thanks also to the radioprotection services of: SCK$\cdot$CEN,
IRMM, University of T\"{u}bingen, MPP M\"{u}nchen and MPIK Heidelberg.
Special thanks to the mechanical workshops of INFN Padova,
MPIK Heidelberg, T\"{u}bingen and IRMM for their commitment
during the production and assembly of the setups.
Finally, we thank all the members of the GERDA collaboration for their warm
support and fruitful discussions.

The project has been supported financially by
the German Federal Ministry for Education and Research (BMBF),
the German Research Foundation (DFG) via the Excellence Cluster,
the Italian Istituto Nazionale di Fisica Nucleare (INFN),
the Max Planck Society (MPG), the Swiss National Science Foundation (SNF).
The institutions acknowledge also internal financial support.


\begin{thebibliography}{9}
\bibitem{bib:comp:vnc}
Virtual Network Computing, a graphical desktop sharing system that uses
the RFB protocol (remote framebuffer) to remotely control another computer.
\end{thebibliography}
\end{document}